\documentclass[%
 aip,
 amsmath,amssymb,
 reprint,%
]{revtex4-1}

\usepackage{graphicx}
\usepackage{dcolumn}
\usepackage{bm}

\usepackage[utf8]{inputenc}
\usepackage[T1]{fontenc}
\usepackage{mathptmx}
\usepackage{etoolbox}
\usepackage{lipsum}
\usepackage{color}
\usepackage{bm}

\makeatletter
\def\@email#1#2{%
 \endgroup
 \patchcmd{\titleblock@produce}
  {\frontmatter@RRAPformat}
  {\frontmatter@RRAPformat{\produce@RRAP{*#1\href{mailto:#2}{#2}}}\frontmatter@RRAPformat}
  {}{}
}%
\makeatother
\begin{document}

\preprint{AIP/123-QED}

\title{Photon scattering from a quantum acoustically modulated two-level system}
\author{Thilo Hahn}
\affiliation{Institute of Solid State Theory, University of M\"unster, 48149 M\"unster, Germany}
\affiliation{Department of Theoretical Physics, Wroc\l{}aw University of Science and Technology, 50-370~Wroc\l{}aw, Poland}

\author{Daniel~Groll}
\affiliation{Institute of Solid State Theory, University of M\"unster, 48149 M\"unster, Germany}
\affiliation{Department of Theoretical Physics, Wroc\l{}aw University of Science and Technology, 50-370~Wroc\l{}aw, Poland}

\author{Hubert J. Krenner}
\affiliation{Institute of Physics, University of M\"unster, 48149 M\"unster, Germany}

\author{Tilmann~Kuhn}
\affiliation{Institute of Solid State Theory, University of M\"unster, 48149 M\"unster, Germany}

\author{Pawe\l{} Machnikowski}
\affiliation{Department of Theoretical Physics, Wroc\l{}aw University of Science and Technology, 50-370~Wroc\l{}aw, Poland}

\author{Daniel~Wigger}
\email{daniel.wigger@pwr.edu.pl}
\affiliation{Department of Theoretical Physics, Wroc\l{}aw University of Science and Technology, 50-370~Wroc\l{}aw, Poland}

\date{\today}

\begin{abstract}
We calculate the resonance fluorescence signal of a two-level system coupled to a quantized phonon mode. By treating the phonons in the independent boson model and not performing any approximations in their description, we also have access to the state evolution of the phonons. We confirm the validity of our model by simulating the limit of an initial quasi-classical coherent phonon state, which can be compared to experimentally confirmed results in the semiclassical limit. In addition we predict photon scattering spectra in the limit of purely quantum mechanical phonon states by approaching the phononic vacuum. Our method further allows us to simulate the impact of the light scattering process on the phonon state by calculating Wigner functions. We show that the phonon mode is brought into characteristic quantum states by the optical excitation process.
\end{abstract}

\maketitle

\section{Introduction}
The development of quantum technologies based on an isolated approach is challenging. Here, hybrid quantum systems offer more flexibility because they combine complementary strengths of dissimilar single systems, while at the same time avoiding their individual shortcomings. The realization of hybrid quantum systems requires different concepts to prepare, store, manipulate, and transmit quantum properties.\cite{kurizki2015quan} In the development of quantum computers, superconducting quantum processors have seen tremendous advances in the last years,\cite{arute2019quantum} while for quantum communication, semiconductor quantum dots (QDs)~\cite{senellart2017high} are state-of-the-art owing to their superior performance with high single-photon emission rates, photon indistinguishability and compatibility with existing fiber optical quantum communication.\cite{xiang2020tune} Very recently, phonons shifted into the focus of hybrid systems and are now considered as an indispensable resource for on-chip hybrid quantum transduction and transfer.\cite{stannigel2010opto, lemonde2018phon} This is rooted in the phonons’ unique, inherent property of coupling to almost any other solid state excitation, including superconducting qubits and single QDs. In addition, the speed of sound in solids is approximately five orders of magnitude lower than the speed of light. Hence, GHz phonons travel with wavelengths in the sub-micrometer range, while the respective optical wavelengths are on the order of centimeters. This promotes phononic circuits for fully-fledged nanoscale integration on quantum chips.

It has been demonstrated that surface acoustic waves (SAWs) can be interfaced with single semiconductor QDs in order to coherently control the single photon emission properties of the QD.\cite{metcalfe2010resolved} We have recently demonstrated that nonlinear wave mixing can be used to precisely tailor the optical scattering spectrum~\cite{weiss2021opto} and that the photon emission time depends on the SAWs' properties.\cite{wigger2021reso} While these recent results were obtained in the semiclassical limit of the phonon coupling, where the phonon field enters purely classically, in the present work we go a crucial step further and model the phonon impact on the quantum level. This renders an important step towards a fully fledged hybrid approach interfacing single solid state quantum systems, single photons, and single phonons. This progress is further promoted by the implementation of acoustic resonator structures.\cite{shao2019phononic,cady2019diamond,nysten2020hybrid,vogele2020quantum} Especially the recent developments in quantum acoustics based on superconductor qubit driven SAW resonators that allows to prepare arbitrary quantum states in an acoustic mode\cite{satzinger2018quan} motivates to investigate the photon-emitter-phonon interface.

In this work we model the mechanical system as a single discrete phonon mode. In many systems one can naturally isolate the coupling to a single phonon mode. For example in the case of longitudinal optical phonons, which typically have a flat dispersion and are energetically well separated from all acoustic modes, the exciton-phonon  coupling of QDs is well described by considering a single mode, while all other modes are not coupled to the exciton.\cite{stauber2000electron} For color centers local lattice vibrations can take this role of the discrete mode.~\cite{gali2011ab} But single frequency phonon fields can also be engineered, e.g., in the form of SAWs\cite{schulein2015four} or in mechanical resonators.\cite{munsch2017reso} Therefore, the retrieved findings are not limited to acoustic resonators but also apply to various other phononic oscillators like optomechanical systems.\cite{cleland2013foun,aspelmeyer2014cavi,bowen2015quantum,kettler2021inducing}

\section{Model}

\begin{figure}[!b]
\includegraphics[width = 0.8\columnwidth]{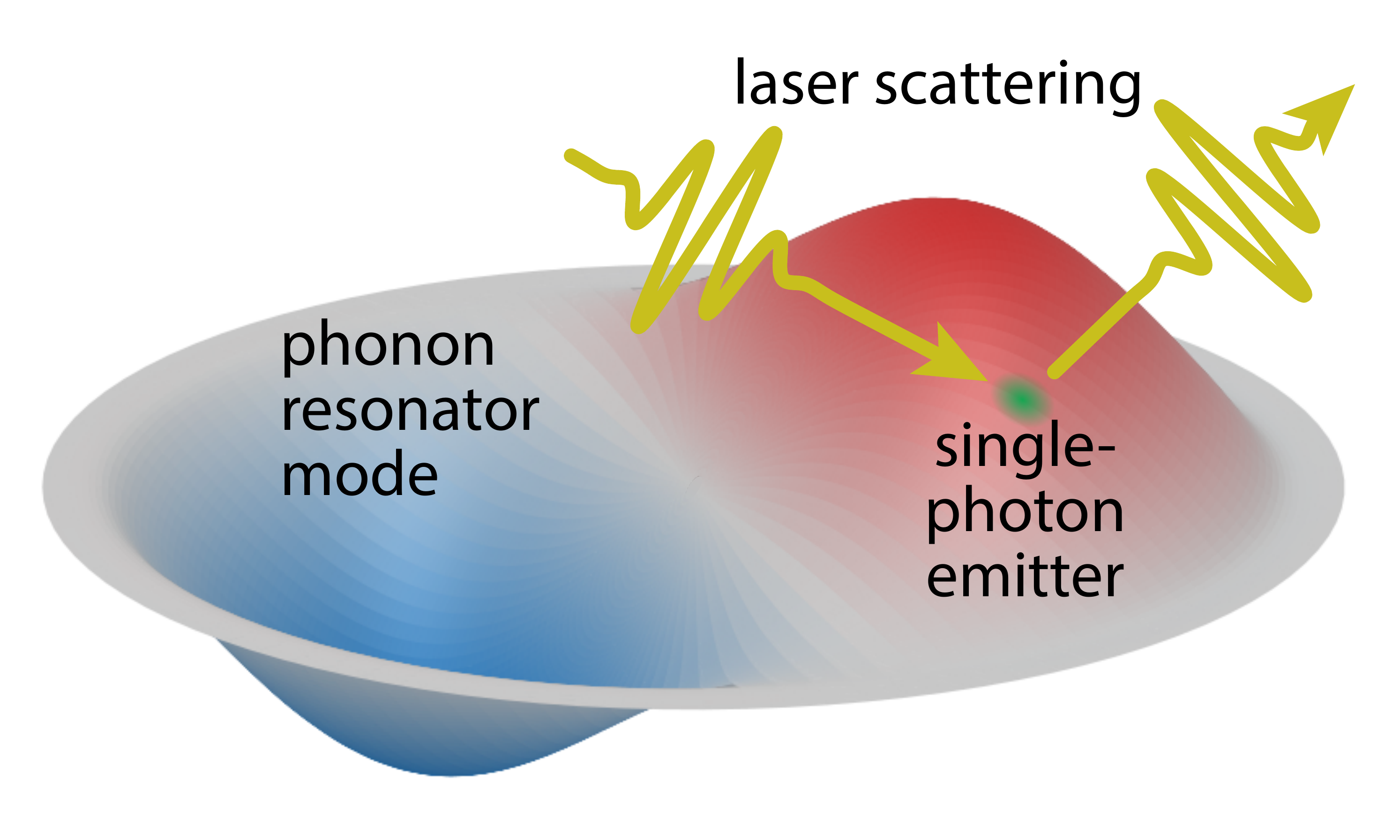}
\caption{Schematic picture of the considered system consisting of a single-photon emitter coupled to a discrete phonon resonator mode. The emitter is excited with laser light and the properties of the scattered light are simulated.}\label{fig:scheme}
\end{figure}

As schematically shown in Fig.~\ref{fig:scheme} we consider a two-level system (TLS) representing a single-photon emitter, which is on the one hand driven by a weak classical laser source and on the other hand coupled to a phononic resonator. Motivated by acoustic and mechanical resonator systems, we restrict ourselves to a single discrete phonon mode. We are considering the transition energy of the TLS to be in the visible range and the phonon energy to be much smaller than this. Therefore, transitions between the ground and excited state of the TLS are exclusively mediated by the optical field. Thus the light-TLS-phonon system is adequately described by the standard single-mode independent boson Hamiltonian with optical driving\cite{mahan2013many,krummheuer2002theory}
\begin{align}\label{eq:H}
H =& \hbar \Omega \left| x \right>\! \left< x \right| + \hbar\omega_\text{ph} b^\dagger b + \hbar g_{\rm ph} \left(b^\dagger + b\right) \left| x \right>\! \left< x \right|  \notag\\
&- \big[ {\bm M} \cdot {\bm E}(t) \left| x \right>\! \left< g \right| + {\bm M}^* \cdot {\bm E}^*(t) \left| g \right>\!\left< x  \right| \big]\,,
\end{align}
where $\left| g \right>$ and $\left| x \right>$ are the TLS ground and excited states, respectively, and $\hbar \Omega$ is the transition energy. The boson operators $b$ and $b^\dagger$ act on the phonon mode with energy $\hbar \omega_\text{ph}$. The TLS-phonon coupling is incorporated by the pure dephasing Hamiltonian with the coupling constant $g_{\rm ph}$. Optical transitions are mediated by the dipole matrix element ${\bm M}$ and the classical optical field ${\bm E}(t)$. This model is well established and was successfully applied to describe different systems from QDs coupled to LO phonons~\cite{castella1999coherent, wilson2002quantum, wilson2004lase, hohenester2009phonon, carmele2010antibunching, kabuss2011microscopic, kabuss2012optically} to (opto)mechanical resonators\cite{li2011nanometer, kanari2021can} or molecules.\cite{gilmore2005spin}

To describe the dynamics of the entire system without approximations, it is convenient to introduce generating functions to express the density matrix of the entire system $\rho$ via\cite{axt1999cohe}
\begin{subequations}\label{eq:generating}
\begin{align}
Y(\alpha,t) &= {\rm Tr}\left[ \vert g \rangle \langle x \vert e^{ -\alpha^* b^\dagger}e^{ \alpha   b} \rho(t) \right]\,, \\
C(\alpha,t) &={\rm Tr}\left[ \vert x \rangle \langle x \vert e^{-\alpha^* b^\dagger}e^{ \alpha b} \rho(t) \right]\label{eq:def_C}\,, \\
G(\alpha,t) &={\rm Tr}\left[ \vert g \rangle \langle g \vert e^{-\alpha^* b^\dagger}e^{ \alpha b} \rho(t) \right] \,. \label{eq:Def_G}
\end{align}
\end{subequations}
While $Y$ describes all phonon-assisted orders of the microscopic polarization of the TLS, $C$ and $G$ describe the occupation of the excited and the ground state of the TLS and the phonons associated with these states, respectively. In the following we will refer to contributions of the system's density matrices of the form $\left<x\right|\rho\left| x\right>$ ($\left<g\right|\rho\left| g\right>$) as phonon state in $\left|x\right>$~($\left|g\right>$).

The Heisenberg equations describing the dynamics of the generating functions read \cite{axt1999cohe}
\begin{subequations}\label{eq:Heis}
\begin{align}
	i\dot Y&(\alpha,t) = \left(\Omega- \frac{i\Gamma}2\right) Y(\alpha,t) \notag\\
			 & +  \big[ \omega_\text{ph}(\alpha \partial_{\alpha} - \alpha^* \partial_{\alpha^*}) + g_{\rm ph}(\alpha - \partial_{\alpha^*} + \partial_{\alpha}) \big] Y(\alpha,t) \notag \\ 
			&+ \frac1\hbar \big[C(\alpha,t)-G(\alpha,t) \big] {\bm M} \cdot {\bm E}(t)\label{eq:Heisa}  \,,\\
	i\dot C&(\alpha,t) = \big[ \omega_\text{ph}(\alpha \partial_{\alpha} - \alpha^*\partial_{\alpha^*}) + g_{\rm ph}(\alpha+\alpha^*)\big] C(\alpha,t) \notag\\
			 &+  \frac1\hbar{\bm M}^* \cdot {\bm E}^*(t)Y(\alpha,t)- \frac1\hbar{\bm M} \cdot {\bm E}(t)Y^*(-\alpha,t)\notag\\
			 &- i\Gamma C(\alpha,t) \label{eq:Heisb} \,,\\
	i\dot G&(\alpha,t) = \omega_\text{ph} (\alpha \partial_\alpha - \alpha^* \partial_{\alpha^*})G(\alpha,t) +i\Gamma C(\alpha,t)\notag\\
			&-\frac1\hbar \big[{\bm M}^* \cdot {\bm E}^*(t) Y(\alpha,t) - {\bm M} \cdot {\bm E}(t)Y^*(-\alpha,t) \big]\,, \label{eq:Heisc} 
\end{align}
\end{subequations}
which render a closed set of equations. In Eqs.~\eqref{eq:Heisb} and \eqref{eq:Heisc} we have additionally included the Markovian dynamics of a radiative decay of the TLS from the excited state into the ground state with the rate $\Gamma$ which is accompanied by a dephasing with the rate $\Gamma/2$ in Eq.~\eqref{eq:Heisa}.\cite{hahn2019infl} To eliminate the partial derivatives with respect to $\alpha$, one introduces transformed fields $\overline Y, \overline C, \overline G$ defined by the characteristics of the Heisenberg equations
\begin{subequations}\label{eq:EOM}
\begin{align}
	{\overline Y} (\alpha, t) &= \exp\left(i{\overline \Omega} t + \gamma\alpha e^{i\omega_\text{ph} t}\right) Y\left(\alpha e^{i\omega_\text{ph} t} - \gamma, t\right) \label{eq:tY}\,,\\ 
	{\overline C} (\alpha, t) &= \exp\left[ 2i\text{Im}\left(\gamma\alpha e^{i\omega_\text{ph} t}\right)\right] C(\alpha e^{i\omega_\text{ph} t}, t)\,,\\
	{\overline G}(\alpha, t) &= G\left(\alpha e^{i\omega_\text{ph} t}, t\right)\label{eq:tG}\,.
\end{align}
\end{subequations}
With these transformations we also introduce the polaron-shifted transition frequency $\overline \Omega = \Omega - g_{\rm ph}^2/\omega_\text{ph}$ of the TLS, the effective phonon coupling strength $\gamma = g_{\rm ph}/\omega_\text{ph}$, and the optical field in the co-rotating frame 
\begin{align}
	\overline E (t)= \frac 1\hbar  {\bm M} \cdot {\bm E}(t) e^{i{\overline \Omega} t}\,.
\end{align}
The equations of motion for the transformed fields read
\begin{subequations}\label{eq:EOM_}
\begin{align}
	i\dot{\overline Y}(\alpha,t) &= {\overline E}(t) \Big[  e^{ \gamma   \alpha  ^*e^{-i\omega_\text{ph}   t}}{\overline  C} \left(\alpha   - \gamma   e^{-i\omega_\text{ph} t} ,t\right) \notag\\
			& \quad - e^{ \gamma   \alpha   e^{i\omega_\text{ph}   t}} {\overline G}\left(\alpha   - \gamma   e^{-i\omega_\text{ph}   t},t\right)\Big]  - \frac {i\Gamma}2 {\overline Y}(\alpha,t)\label{eq:Y}\,,\\
	i\dot{\overline C}(\alpha,t) &= e^{-\gamma^2}\Big\lbrack e^{-\gamma\alpha^* e^{-i\omega_\text{ph} t}} {\overline E}^*(t){\overline Y}\left(\alpha + \gamma e^{-i\omega_\text{ph} t},t\right)\notag\\
		&\quad - e^{\gamma \alpha e^{i\omega_\text{ph} t}}{\overline E}(t){\overline Y}^*\left(-\alpha + \gamma e^{-i\omega_\text{ph} t}\right)\Big\rbrack  - i\Gamma\overline{C}(\alpha, t)\,,\label{eq:C}\\
i\dot{\overline G}(\alpha,t) &=e^{-\gamma^2}\Big\lbrack -e^{-\gamma\alpha e^{i\omega_\text{ph} t}} {\overline E}^*(t){\overline Y}\left(\alpha + \gamma e^{-i\omega_\text{ph} t},t\right)\notag\\
		&\quad + e^{\gamma \alpha^* e^{-i\omega_\text{ph} t}}{\overline E}(t){\overline Y}^*\left(-\alpha + \gamma e^{-i\omega_\text{ph} t}\right)\Big\rbrack \notag\\
&\quad + i\Gamma e^{-2i\gamma\text{Im}\left(\alpha e^{i\omega_\text{ph} t}\right)} \overline{C}\left(\alpha ,t\right)\,. \label{eq:G}
\end{align}
\end{subequations}
Although they do not contain partial derivatives with respect to $\alpha$ anymore, now the time-evolution of a generating function at a given point $\alpha$ depends on the other functions at different points when the optical field $\overline{E}$ is driving the system. In the absence of optical driving $\overline{E}=0$ Eqs.~\eqref{eq:EOM_} can directly be solved. To recover the original generating functions, the back-transformations read
\begin{subequations}
\begin{align}
	Y(\alpha, t) &= e^{-i\overline\Omega t - \gamma(\alpha + \gamma)}\overline Y \left[e^{-i\omega_\text{ph} t}( \alpha + \gamma), t\right]\label{eq:bY}\,,\\
	C(\alpha, t) &= e^{-2i\text{Im}(\gamma\alpha)}\overline C\left(e^{-i\omega_\text{ph} t}\alpha, t\right)\label{eq:bC}\,,\\
	G(\alpha, t) &=  \overline G\left(e^{-i\omega_\text{ph} t}\alpha, t\right)\,.\label{eq:bG}
\end{align}
\end{subequations}

\section{Expansion for weak optical driving}
For arbitrary fields the dynamics of the generating functions have to be solved numerically,\cite{axt1999cohe} while for ultrashort pulses exact solutions are known.\cite{vagov2002elec} We are here interested in a weak continuous optical driving with $\overline{E}(t)=E_0 e^{-i(E_{\rm laser}/\hbar - \overline{\Omega})t}$. Our approach to approximate this situation is to expand the equations in powers of the light field to retrieve the optical signal in the second order of $\overline{E}$ which is sufficient to simulate the optical scattering spectrum.\cite{weiss2021opto} Therefore, we express the generating functions into a series $Y = Y^{(0)} + Y^{(1)} + Y^{(2)}+\dots$, where $Y^{(i)} \sim \mathcal{O}(E^i)$ and analogously for $C$ and $G$. Corresponding expansions hold for the transformed generating functions $\overline Y$, $\overline C$, and $\overline G$. When initially starting in the ground state $\left|g\right>$ at $t_0$ with a given $G^{(0)}(t_0)$ and the corresponding $\overline{G}^{(0)}(t_0)$, in the zeroth order, there is no polarization or occupation
\begin{align}
	Y^{(0)}(t_0) = {\overline Y}^{(0)}(t_0) = C^{(0)}(t_0) = {\overline C}^{(0)}(t_0) = 0 \,.
\end{align}

From the initial $\overline{G}^{(0)}(t_0)$ we obtain the transformed polarization function in the first order of the light field $\overline{Y}^{(1)}$ by formally integrating Eq.~\eqref{eq:Y}. Then we retrieve the polarization function by applying the back-transformation in Eq.~\eqref{eq:bY}:
\begin{align}\label{eq:Y1}
	&Y^{(1)}(\alpha,t) = i e^{-i\overline \Omega t}\int\limits_{t_0}^{\ t} dt' \exp\left\{ \gamma(\alpha+\gamma)\left[e^{i\omega_{\rm ph}(t'-t)}-1\right]\right\}  \notag\\
		& \times\overline G^{(0)}\left[ e^{-i\omega_{\rm ph} t}(\alpha+\gamma)-\gamma e^{-i\omega_{\rm ph} t'}, t_0\right] e^{- \frac{\Gamma}2 (t-t')} \overline E\left(t'\right)\,,
\end{align}
where we have used that $\overline{G}^{(0)}$ is time-independent according to Eq.~\eqref{eq:G}.

To gain access to the phonon state in the excited state of the TLS $\left|x\right>$ and to calculate the optical scattering spectrum we have to calculate the occupation function $C$. Looking at Eqs.~\eqref{eq:Heis} we see that the occupation has no contribution in the linear order of the optical field $C^{(1)}=0$. We calculate its first non-vanishing contribution by integrating Eq.~\eqref{eq:C}, inserting $\overline{Y}^{(1)}$ and transforming it back with Eq.~\eqref{eq:bC}. With this procedure we retrieve
\begin{align}\label{eq:clong}
	 &C^{(2)}(\alpha,t) =\notag\\
		 & \int\limits_{t_0}^{\ t} dt' \exp\left\{ \gamma\alpha \left[e^{i\omega_{\rm ph} (t'-t)}-1 \right] - \gamma\alpha^* \left[e^{-i\omega_{\rm ph} (t'-t)}-1\right] \right\}  \notag\\
		& \times e^{-\Gamma(t-t')} \Bigg( \int\limits_{t_0}^{\ t'}dt'' \ \overline E^*(t') \overline E(t'') e^{-\frac\Gamma2 (t'-t'')}\notag\\
		&\qquad \times \exp\left\{ \gamma \left[\alpha e^{i\omega_{\rm ph}(t'-t)} + \gamma\right]\left[e^{i\omega_{\rm ph} (t''-t')}-1\right] \right\} \notag\\
		&\qquad\times\overline G^{(0)} \left\{ e^{-i\omega_{\rm ph} t'}\left[\alpha e^{i\omega_{\rm ph}(t'-t)} + \gamma \right] - \gamma e^{-i\omega_{\rm ph} t''}, t_0 \right\}  \notag\\
		&\qquad\qquad +\int\limits_{t_0}^{\ t'} dt''\ \overline E^*(t'') \overline E(t')e^{-\frac\Gamma2 (t'-t'')} \notag\\
		&\qquad\times \exp\left\{ \gamma\left[-\alpha^* e^{-i\omega_{\rm ph}(t'-t)} + \gamma \right]\left[e^{-i\omega_{\rm ph} (t''-t')}-1\right] \right\} \notag\\
		&\qquad\times\overline G^{(0)}\left\{ e^{-i\omega_{\rm ph} t'}\left[\alpha e^{i\omega_{\rm ph}(t'-t)} - \gamma \right] + \gamma e^{-i\omega_{\rm ph} t''}, t_0\right\} \Bigg)\,,
\end{align}
where we have used that $G^*(\alpha,t) = G(-\alpha,t)$ which can be seen from conjugating the definition of the generating function in Eq.~\eqref{eq:Def_G}. Note, that after exchanging $t'\leftrightarrow t''$ in the last double-integral, Eq.~\eqref{eq:clong} contains both time-orderings $t'>t''$ and $t''>t'$ in a symmetric way. Thus, we can combine the two integrals by expanding the $t''$ integral to $t$
\begin{align}\label{eq:c2}
		&C^{(2)} (\alpha,t)= e^{-\gamma (\alpha-\alpha^* +\gamma)} \iint\limits_{t_0}^{\quad \ t} dt' dt''\ \overline E^*(t') \overline E(t'') e^{- \frac{\Gamma}{2} (2t-t'-t'')} \notag\\
		&\times  \exp\left\{ \gamma\left[\alpha e^{i\omega_\text{ph}(t''-t)} - \alpha^* e^{-i\omega_\text{ph} (t'-t)} + \gamma e^{i\omega_\text{ph} (t''-t')} \right]\right\}  \notag\\
		& \times  \overline G^{(0)}\left( \alpha e^{-i\omega_\text{ph} t} + \gamma e^{-i\omega_\text{ph} t'} - \gamma e^{-i\omega_\text{ph} t''},t_0 \right) \,.
\end{align}

An instructive quantity to study the phonon's quantum state is the Wigner function. For the excited state, it can be directly constructed from the occupation function via the Fourier transform\cite{wigger2016quan}
\begin{align}\label{eq:Wigner}
	&W_x(U, \Pi;t) = \notag\\
	&\frac 1{4\pi} \iint\limits_{-\infty}^{\quad\infty}d^2\alpha\, C(\alpha,t) \exp\left\{-\frac{\vert \alpha \vert ^2}2-i\big[\Pi\text{Re} (\alpha) + U\text{Im}(\alpha)\big]\right\} \,. 
\end{align}
The phase space hosting the Wigner function is spanned by the two variables $U$ and $\Pi$ that represent the displacement $u=b+b^\dagger$ and momentum $p=(b-b^\dagger)/i$ of the phonon mode, respectively. The operators are connected to the phase space variables via the eigenvalue equations $u\left|U\right>=U\left|U\right>$ and $p\left|\Pi\right>=\Pi\left|\Pi\right>$. The Wigner function for the phonon state in the ground state of the TLS $W_g$ is calculated by replacing $C(\alpha)$ by $G(\alpha)$ in Eq.~\eqref{eq:Wigner}. Consequently the Wigner function of the full phonon state is given by $W=W_g+W_x$. It is important to note here, that the equilibrium position (fixed point of the phase space dynamics) for $W_g$ is at $(U,\Pi)=(0,0)$, while the equilibrium for $W_x$ is shifted to $(U,\Pi)=(-2\gamma,0)$.\cite{groll2021controlling}

\section{Optical signal}

\begin{figure}[t]
\centering
\includegraphics[width = 0.8\columnwidth]{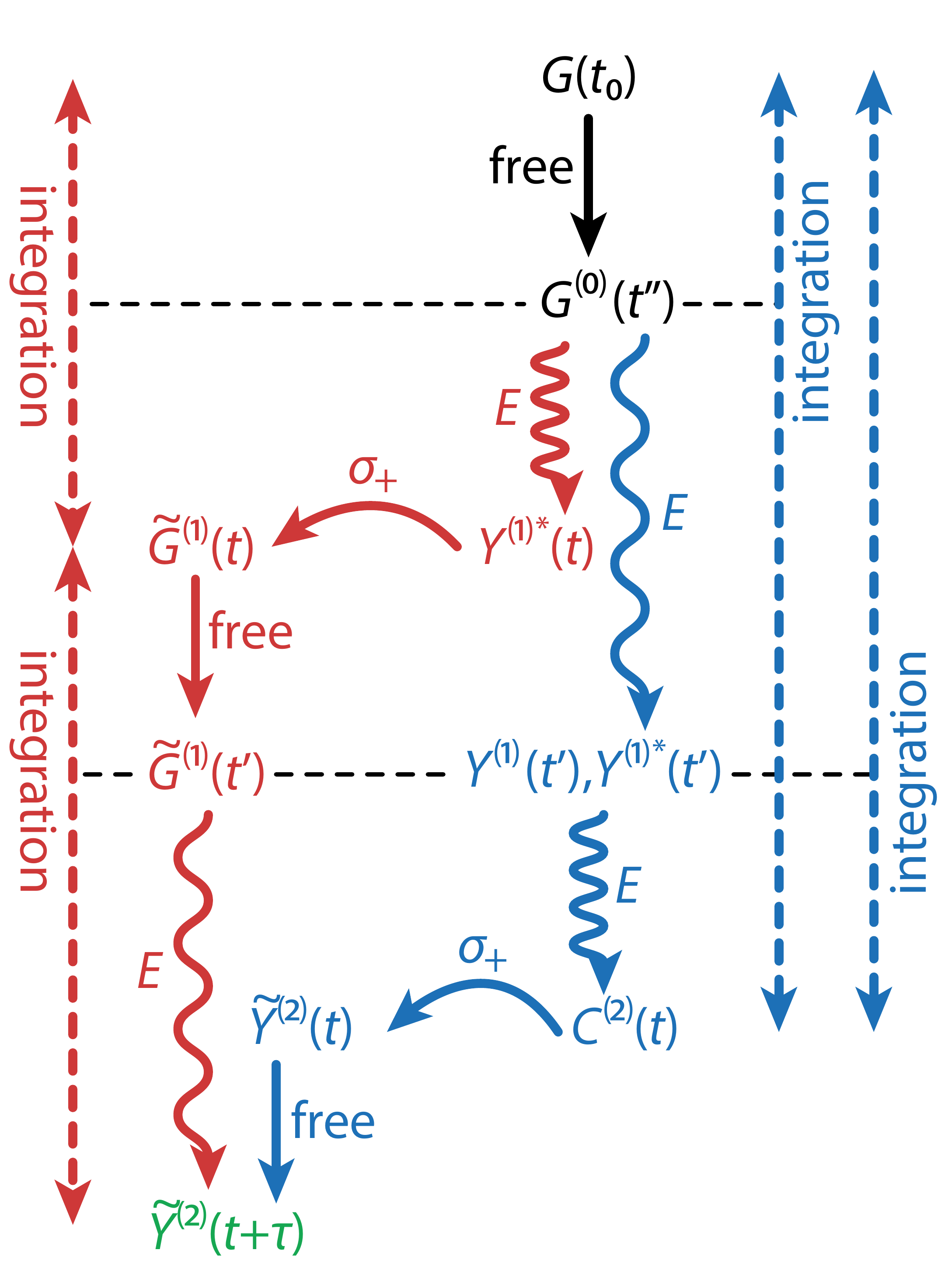}
\caption{Schematic picture of the calculation of the two-time correlation function. The first path is shown in red (left) and the second one in blue (right). Straight solid arrows show a field-free propagation. Waved arrows refer to optically induced transitions. Dashed arrows indicate which quantities are integrated over which intervals.}\label{fig:scheme_RF}
\end{figure}
The optical scattering spectrum of a single-photon emitter is given by\cite{weiss2021opto}
\begin{align}
	S(t,\omega) = 2\text{Re}\left[\int\limits_{0}^\infty d\tau\ e^{i\omega \tau-\eta \tau} \mathcal G^{(1)}(t,t+\tau)\right]\,,
\end{align}
where we have included a line-broadening $\eta$ to account for a non-vanishing detection resolution.\cite{groll2021controlling} The two-time correlation function $\mathcal G^{(1)}(t,t+\tau)$ of the TLS includes the raising $\sigma_+ = \vert x \rangle \langle g \vert$ and lowering $\sigma_- = \vert g \rangle \langle x \vert$ operators via
\begin{align}
	\mathcal G^{(1)}(t,t+\tau) = \langle \sigma_+(t) \sigma_-(t+\tau)\rangle\,.
\end{align}
The corresponding time integrated spectrum is given by
\begin{align}
	\overline{S}(\omega)= \frac{1}{T} \int\limits_{T_0}^{T_0+T} dt\ S(t,\omega) \,,\label{eq:S_int}
\end{align}
where we assume that $\mathcal G^{(1)}$ is periodic in $t$ and $\tau$ with the period $T$. Here, the dynamics of the entire system including TLS and phonons is Markovian and the correlation function can be calculated by using the quantum regression theorem~\cite{meystre2007}
\begin{align}\label{eq:G_liouville}
	\mathcal G^{(1)}(t,t+\tau) = \text{Tr}\Big\{ \mathcal L (t+\tau, t) \big[\rho(t) \sigma_+  \big] \sigma_-  \Big\}\,,
\end{align}
where $\mathcal L(t+\tau,t)$ is the evolution super-operator that propagates the density matrix $\rho$ from $t$ to $t+\tau$. We can write the density matrix in the form
\begin{align}\label{eq:phonon_rho}
	\rho = \left| g\right>\!\left< g\right| \rho_{\rm ph}^{gg} + \left| x\right>\!\left< x\right| \rho_{\rm ph}^{xx} + \left( \left| g\right>\!\left< x\right| \rho_{\rm ph}^{gx} + h.c.\right)\,,
\end{align}
where the effective phonon density matrices $\rho_{\rm ph}^{ij}=\left<i\right|\rho\left|j\right>$ describe the phonon states associated with the occupations and the polarization. From Eqs.~\eqref{eq:generating} we see that there is a one-to-one correspondence between the generating functions and the effective phonon density matrices of the form 
\begin{subequations}\label{eq:generating_2}
	\begin{align}
	Y(\alpha,t) &= {\rm Tr}\left[ e^{ -\alpha^* b^\dagger}e^{ \alpha   b} \rho_{\rm ph}^{xg}(t) \right]\,, \\
	C(\alpha,t) &={\rm Tr}\left[ e^{-\alpha^* b^\dagger}e^{ \alpha b} \rho_{\rm ph}^{xx}(t) \right]\label{eq:def_C_2}\,, \\
	G(\alpha,t) &={\rm Tr}\left[ e^{-\alpha^* b^\dagger}e^{ \alpha b} \rho_{\rm ph}^{gg}(t) \right] \,. \label{eq:Def_G_2}
	\end{align}
\end{subequations}
The action of the operator $\sigma_+$ at time $t$ in Eq.~\eqref{eq:G_liouville} basically relabels the effective phonon density matrices, as can be seen by evaluating
\begin{align}
	\widetilde{\rho}=\rho\sigma_+&=\left| x\right>\!\left< g\right| \rho_{\rm ph}^{xx} + \left| g\right>\!\left< g\right| \rho_{\rm ph}^{gx}\notag\\
		&=\left| x\right>\!\left< g\right| \widetilde{\rho}_{\rm ph}^{xg} + \left| g\right>\!\left< g\right| \widetilde{\rho}_{\rm ph}^{gg}\,.
\end{align}
$\widetilde{\rho}$ is then propagated between times $t$ and $t+\tau$. To simulate this with generating functions, we have to determine them based on $\widetilde{\rho}$, where the only two non-vanishing contributions are
\begin{subequations}\label{eq:generating_3}
	\begin{align}
	\widetilde{Y}(\alpha,t) &= {\rm Tr}\left[ \vert g \rangle \langle x \vert e^{ -\alpha^* b^\dagger}e^{ \alpha   b} \widetilde{\rho}(t) \right]=C(\alpha,t)\,, \label{eq:Ytilde}\\
	\widetilde{G}(\alpha,t) &={\rm Tr}\left[ \vert g \rangle \langle g \vert e^{-\alpha^* b^\dagger}e^{ \alpha b} \widetilde{\rho}(t) \right]=Y^*(-\alpha,t) \,. \label{eq:Gtilde}
	\end{align}
\end{subequations}
These two relations render initial conditions for the propagation from time $t$ to $t+\tau$. For that purpose the generating functions $\widetilde{Y}$ and $\widetilde{G}$ are also propagated as $Y$ and $G$, respectively, via the Heisenberg equations Eqs.~\eqref{eq:Heis} in the first order of the optical field. Therein we need to set ${\widetilde Y}^* (\alpha,t) = 0$  to take into account that $\widetilde \rho(t)$ is not Hermitian.

As a consequence the quantum regression theorem is applicable here when utilizing generating functions. The individual steps to calculate the optical spectrum are depicted in Fig.~\ref{fig:scheme_RF}. The two possible depicted paths resemble the two non-vanishing density matrix elements after the application of the TLS raising operator from Eqs.~\eqref{eq:generating_3}.

The procedure to derive the correlation function follows the same strategy as used in Ref.~[\onlinecite{weiss2021opto}] (see Supporting Material therein). Starting from the ground state $G(t_0)=G^{(0)}(t_0)$ and $Y(t_0)=C(t_0)=0$ we first follow the red path, with a light-induced transition into the polarization ${Y^{(1)}}^*(t)$ (Eq.~\eqref{eq:Y1}). After application of the TLS raising operator, it becomes the new ground-state function $\widetilde{G}^{(1)}(t)$ (Eq.~\eqref{eq:Gtilde}). Propagating this function with another light-induced step using once more Eq.~\eqref{eq:Y1} with $G$ replaced by $\widetilde G$, it becomes the final polarization function
\begin{align}
	&\widetilde Y_1^{(2)}(\alpha,t+\tau) =\notag\\
	& ie^{-i\overline\Omega\tau - \gamma(\alpha+\gamma)}\int\limits_0^{\ \tau} dt' \exp\left\{ \gamma\left[ e^{-i\omega_{\rm ph} \tau}(\alpha + \gamma)\right] e^{i\omega_{\rm ph} t'} \right\} \notag\\
			&\qquad \times {Y^{(1)}}^* \left[ -e^{-i\omega_{\rm ph} \tau}(\alpha + \gamma) + \gamma e^{-i\omega_{\rm ph} t'}, t \right] \overline E(t'+t)\,,
\end{align}
which is in second order of the optical field.

Secondly, we follow the blue path. From the ground state function $G^{(0)}(t_0)$ the occupation function $C^{(2)}(t)$ is created by two light-induced transitions via an intermediate polarization ${Y^{(1)}}^{(*)}$ (Eq.~\eqref{eq:c2}). Applying the TLS raising operator gives the polarization function $\widetilde{Y}^{(2)}(t)$ (Eq.~\eqref{eq:Ytilde}). We finally propagate $\widetilde{Y}^{(2)}(t)$ from $t$ to $t+\tau$ by transforming it with Eq.~\eqref{eq:tY}, applying Eq.~\eqref{eq:Y}, and transforming it back with Eq.~\eqref{eq:bY} leading to the second contribution
\begin{align}
	&\widetilde Y_2^{(2)} (\alpha,t+\tau) = \notag\\
		& \exp\left\{ -i\overline\Omega \tau - \gamma(\alpha + \gamma) + \gamma\left[ e^{-i\omega_{\rm ph}\tau}(\alpha + \gamma)\right] \right\} \notag\\
		&\qquad \times C^{(2)}\left[ e^{-i\omega_{\rm ph}\tau}(\alpha + \gamma)-\gamma,t\right]\,.
\end{align}
Adding both results at $\alpha = 0$ gives the correlation function
\begin{align}\label{eq:RF}
	&\mathcal G^{(1)}(t,t+\tau) =  \left.\left[ \widetilde Y_1^{(2)} (\alpha,t+\tau)+ \widetilde Y_2^{(2)} (\alpha,t+\tau) \right]\right|_{\alpha=0} \notag \\
	&=e^{-i\overline\Omega\tau } \int\limits_{t_0}^{\ t} dt' \int\limits_{t_0}^{\ t+\tau}dt''\ \overline E^*(t') \overline E(t'') e^{- \frac{\Gamma}2 (2t+\tau-t'-t'')}\notag\\
			&\quad\times  \exp \left\{ \gamma^2 \left[ e^{-i\omega_{\rm ph}(t+\tau-t'')} + e^{-i\omega_{\rm ph} (t'-t)}  -2 \right.\right.\notag\\
			&\qquad\qquad \left.\left. + e^{i\omega_{\rm ph} (t''-t')} - e^{i\omega_{\rm ph}(t + \tau - t')}    - e^{i\omega_{\rm ph}(t''-t)} + e^{i\omega_{\rm ph}\tau} \right]\right\}   \notag\\
			& \quad\times  \overline G^{(0)} \left[ \gamma\left(e^{-i\omega_{\rm ph}(\tau+t)}- e^{-i\omega_{\rm ph} t} +  e^{-i\omega_{\rm ph} t'} -  e^{-i\omega_{\rm ph} t''} \right), t_0\right]\,.
\end{align}
This expression now allows us to calculate the light scattering spectrum with an arbitrary initial phonon quantum state in the ground state of the TLS, which is included by the specific choice of $G^{(0)}(t_0)$ in Eq.~\eqref{eq:Def_G}.

\section{Results}
\begin{figure}[t]
\centering
\includegraphics[width =0.8 \columnwidth]{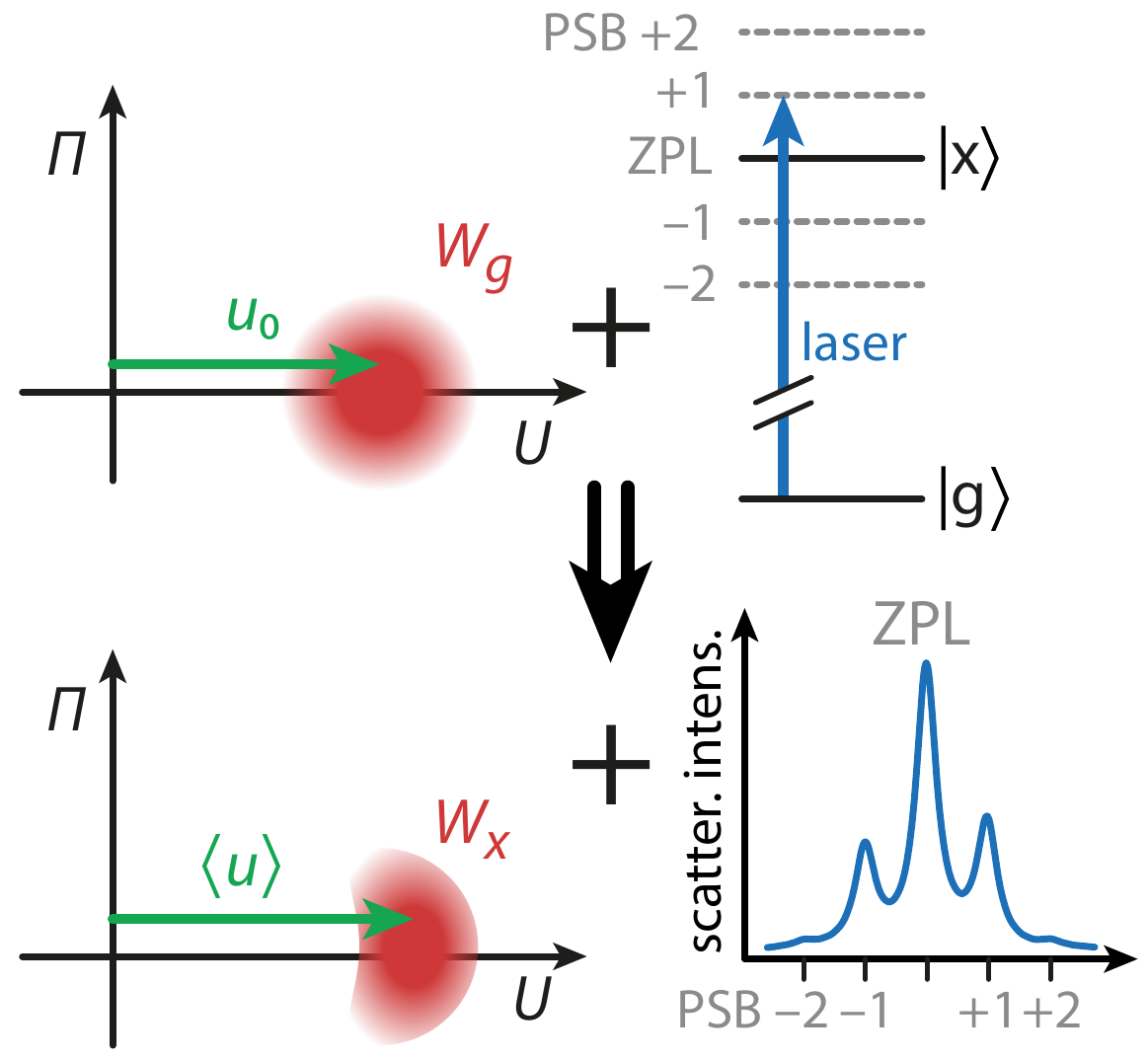}
\caption{Schematic picture of the performed study. The initial condition is depicted in the top, where we consider the phonons being in a coherent state with amplitude $u_0$ and the TLS being in the ground state $\left|g\right>$. In addition we apply a continuous wave laser that is in resonance with the ZPL or one of the PSBs. During the optical driving we study the phonon state in the excited state $\left|x\right>$ and the light scattering spectrum. The phonon states are illustrated by the Wigner functions $W_g$ and $W_x$.}\label{fig:scheme_study}
\end{figure}

To investigate the interplay between an initially prepared phonon quantum state and the light scattered from the TLS we perform the analysis sketched in Fig.~\ref{fig:scheme_study}. For this proof-of-concept study we restrict ourselves to initial coherent phonon states with the displacement $u_0=\left<u\right>(t=t_0)$ in the ground state of the TLS $\left| g\right>$ which is described by the Wigner function\cite{reiter2011gene}
\begin{align}
	W=W_g(U,\Pi) = \frac{1}{2\pi} \exp\left\{-\frac 12 \left[ (U-u_0)^2+\Pi^2\right]\right\}\,.
\end{align}%
Furthermore, we consider different energies for the optical excitation $\overline{E}(t)=E_0 e^{-i\Delta_{\rm laser} t/\hbar}$ that can be in resonance with the TLS transition $\Delta_{\rm laser}=E_{\rm laser}-\hbar\overline{\Omega}=0$, called zero-phonon line (ZPL) or with one of the phonon sidebands (PSBs), i.e., $\Delta_{\rm laser}= n\,\hbar\omega_{\rm ph}$, $n=\pm 1,\,\pm2,\,\dots$ . Our method allows us not only to simulate the light scattering spectrum induced by the phonon coupling, but also the phonon state in the excited state of the TLS $\left|x\right>$ which can be modified by the interplay with the light-driven TLS with respect to the initial state.

Before embarking on the classical-to-quantum transition regime for the phonons we focus on two limiting cases in the following subsections.

\subsection{Semiclassical limit}

Recent studies have focused on light scattering experiments from a single semiconductor QD in the presence of SAW fields.\cite{weiss2021opto, wigger2021reso} There, it was shown that the treatment of the SAW in the semiclassical limit by simply considering a time-dependent transition energy of the QD's leads to excellent agreements with the experiments. As first benchmark for our quantum acoustic model we consider the semiclassical limit of a coherent phonon state with a large amplitude.~\cite{glauber1963coherent}

Using that the coherent state is an eigenstate of the annihilation operator $b$ the initial phonon state translates to the ground state function
\begin{align}
	G^{(0)}(t_0) &= \text{Tr} \left(e^{\alpha b} \left| \frac{u_0}{2}\right>\!\left< \frac{u_0}{2}\right| e^{-\alpha^* b^\dagger}\right) \notag\\
		&= \exp\left[ i u_0\,\text{Im}\left(\alpha \right)\right] \,.
\end{align}
Inserting this phonon state into Eq.~\eqref{eq:RF} leads to the correlation function for an initial coherent state
\begin{align}
	&\mathcal G^{(1)}(t,t+\tau) =  \notag\\
		&e^{-i\overline\Omega\tau } \int\limits_{t_0}^{\ t} dt'\int\limits_{t_0}^{\ t+\tau} dt''\  \overline E^*(t')  \overline E(t'') e^{-\frac\Gamma2(2t+\tau-t'-t'')}   \notag\\
			&\quad\times e^{ i u_0 \gamma  \lbrace \sin(\omega_{\rm ph} t) - \sin(\omega_{\rm ph} t')  + \sin(\omega_{\rm ph} t'') - \sin\lbrack\omega_{\rm ph}(t+\tau)\rbrack\rbrace}\notag\\
			&\quad\times  \exp\left\{ \gamma^2 \left[e^{-i\omega_{\rm ph}(t+\tau-t'')} + e^{-i\omega_{\rm ph} (t'-t)}  -2  \right.\right.\notag\\
			&\qquad \left.\left. + e^{i\omega_{\rm ph} (t''-t')} - e^{i\omega_{\rm ph}(t + \tau - t')}    - e^{i\omega_{\rm ph}(t''-t)} + e^{i\omega_{\rm ph}\tau}\right]\right\}\,.
\end{align}
The semiclassical phonon limit is reached for a weak coupling $\gamma$ and a large initial displacement $u_0$. Therefore, we consider $\gamma\to0$ while keeping $u_0\gamma = D={\rm const.}$, resulting in
\begin{align}\label{eq:classical}
	&\mathcal G^{(1)}(t,t+\tau) =\notag\\
		& e^{-i\overline\Omega\tau } \int\limits_{t_0}^{\ t} dt' \!\!\int\limits_{t_0}^{\ t+\tau} dt''\  \overline E^*(t')  \overline E(t'') e^{-\frac\Gamma2(2t+\tau-t'-t'')}   \notag\\
		&\quad\times e^{ i D  \left\{ \sin(\omega_{\rm ph} t) - \sin(\omega_{\rm ph} t')  + \sin(\omega_{\rm ph} t'') - \sin\lbrack\omega_{\rm ph}(t+\tau)\rbrack\right\}  } \,.
\end{align}
This result nicely agrees with the semiclassical model in Ref.~[\onlinecite{wigger2021reso}], where the parameter $D$ was considered as the relative SAW induced energy shift of the TLS transition energy $\Delta_0/\omega_{\rm SAW}$. Although we have for simplicity chosen a real initial coherent state amplitude $u_0$, the result for the correlation function in the semiclassical limit is also reached for any other initial phase of the coherent state.

\subsection{Initial phonon vacuum state}
For an initial phonon vacuum state $\left|g,0\right>$, the ground state function is simply $\overline G^{(0)}(t_0) = 1$. With this we can express the correlation function as
\begin{align}
	&\mathcal G^{(1)}(t,t+\tau) = \notag\\
		&e^{-i\overline\Omega\tau }  \iint\limits_{-\infty}^{\quad 0} dt'dt''\ \overline E^*(t+t')  \overline E(t+\tau+t'') e^{\frac\Gamma2(t'+t'')}\notag\\
		 &\qquad\times\exp\left[ \gamma^2\left( e^{i\omega_{\rm ph} t''} + e^{-i\omega_{\rm ph} t'}-2\right)\right] \notag\\
		 &\qquad\times\exp\left\{ \gamma^2\left[ e^{i\omega_{\rm ph} \tau}\left(1-e^{i\omega_{\rm ph} t''}\right)\left(1-e^{-i\omega_{\rm ph} t'}\right) \right]\right\}\,,
\end{align}
where we have assumed that $\Gamma(t-t_0)\gg1$ and substituted $t'\to t+t'$ and $t''\to t''+t+\tau$ shifting the lower integral limit to $-\infty$. Expanding the exponential function in the last line, we factorize the double integral into
\begin{align}
	&\mathcal G^{(1)}(t,t+\tau) =  e^{-i\overline\Omega\tau } \sum_n \frac{\gamma^{2n}}{n!} e^{in\omega_{\rm ph} \tau}\notag\\
		&\qquad \times \int\limits_{-\infty}^{\ 0}dt'\ \overline E^*(t+t') e^{\frac\Gamma2 t'} \left(1- e^{-i\omega_{\rm ph} t'}\right)^n \notag \\
			&\qquad\qquad\times \exp\left[-\gamma^2\left(1-e^{-i\omega_{\rm ph} t'}\right)\right]\notag\\
			&\qquad\times\int\limits_{-\infty}^{\ 0}dt''\  \overline E(t+\tau+t'')e^{\frac\Gamma2 t''} \left(1-e^{-i\omega_{\rm ph} t''}\right)^n \notag \\
				&\qquad\qquad\times \exp\left[-\gamma^2\left(1-e^{-i\omega_{\rm ph} t''}\right) \right]  \,.
\end{align}
While this expression looks quite involved, one can at least estimate that the correlation function is peaked for a small decay rate $\Gamma \to 0$ ($t_0\to -\infty$) when the optical field is in resonance with a phonon-assisted transition, i.e., $\Delta_{\rm laser}=\hbar m\omega_{\rm ph}$ or $\overline E(t) = E_0 e^{-im \omega_{\rm ph} t}$ with $m=0,\, 1,\,  2 ,\, \dots$\,. Then, we can elegantly write
\begin{align}\label{eq:RF_vac}
	&\mathcal G_m^{(1)}(t,t+\tau) = e^{-i\overline\Omega\tau } E_0^2\notag\\
	& \left. \times\sum_n \frac{\gamma^{2n}}{n!} e^{i\omega_{\rm ph}(n-m)\tau} \left\lbrack \partial_\lambda^n e^{\lambda} \frac{(-\lambda)^m}{m!} \right\rbrack^2\right|_{\lambda = -\gamma^2}\,.
\end{align}
In the most simple case of a resonant excitation on the polaron-shifted transition energy, i.e., on the ZPL with $m=0$, we obtain
\begin{align}\label{eq:vacuum_limit}
	\mathcal G_0^{(1)}&(t,t+\tau) =e^{-i\overline\Omega\tau }  E_0^2\sum_n e^{-\gamma^2} \frac{\gamma^{2n}}{n!} e^{in\omega_{\rm ph} \tau}\,,
\end{align}
which is the characteristic function of a Poisson distribution with expectation value $\gamma^2$ leading to a Poisson distributed spectrum.

\subsection{Modification of the phonon state}\label{sec:phonons}

To approach the classical-to-quantum transition regime we here study the modification from the initial coherent phonon state by the light scattering from the TLS under the considered CW-excitation starting at $t_0$. If the optical driving is sufficiently weak, the excited state occupation will be dominated by the second order in the optical field on the timescale $ E_0^{-2}$. If there is a decay of the TLS, this manifests in a form-stable Wigner function for $\Gamma(t-t_0)\gg1$ which rotates in phase space with the frequency $\omega_\text{ph}$ around its respective equilibrium position. This can be seen from Eq.~\eqref{eq:c2} by substituting $t'\to t+t'$ and $t''\to t+t''$. In Appendix \ref{sec:stable} we show that it is sufficient to evaluate the Wigner function after a whole phonon period to obtain its form-stable shape for an excitation on the ZPL or on a PSB. 

\begin{figure}[!t]
\includegraphics[width = \columnwidth]{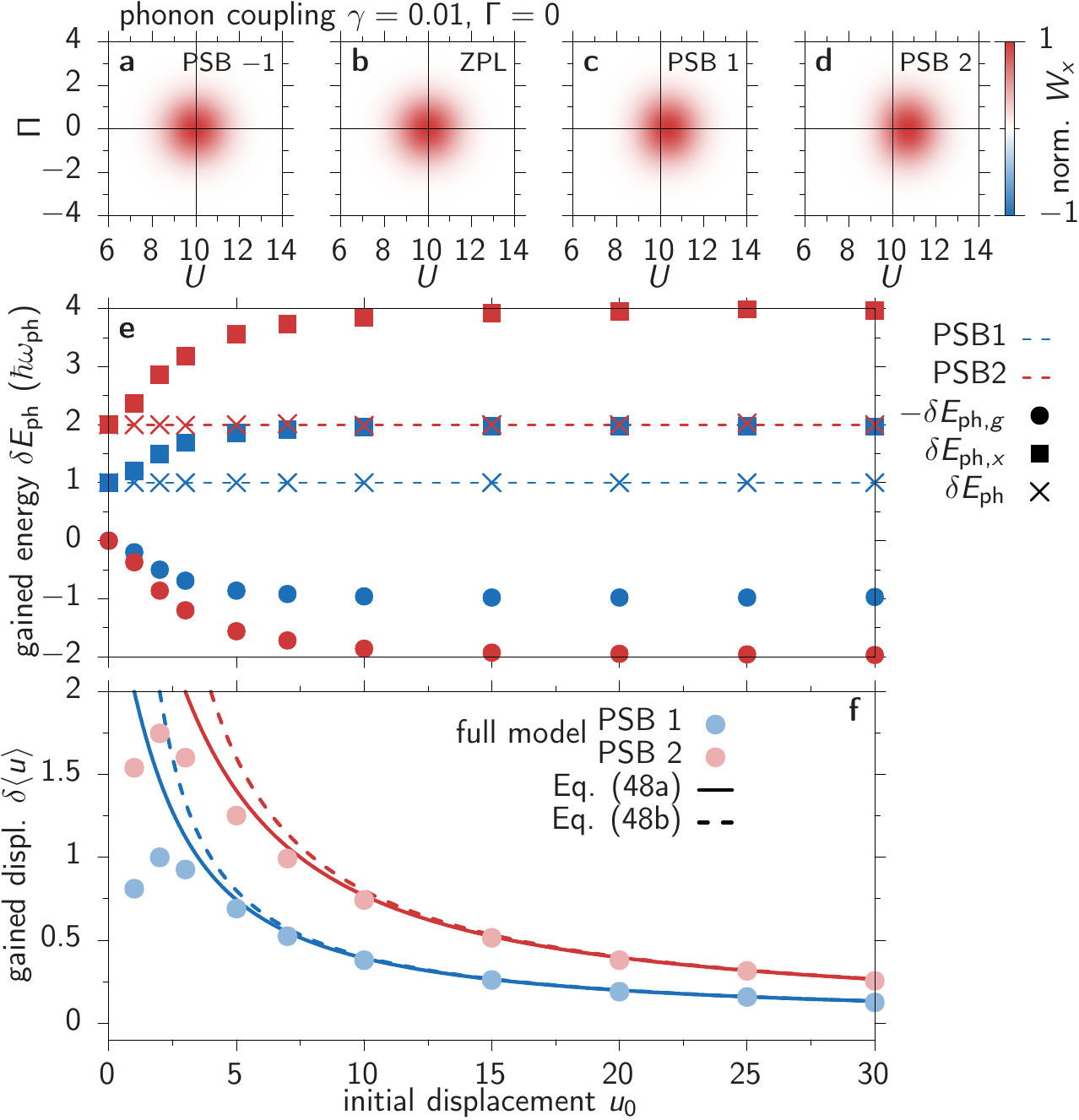}
\caption{(a-d) Normalized Wigner function of the excited state for $u_0=10$ and laser detuning increasing from PSB $-1$ to PSB 2 from left to right. (e) Energy gain of the ground and excited state phonons as dots and squares. Sum of energy changes marked as crosses.  Laser detuning on the PSB 1 ($\Delta_{\rm laser}=\hbar\omega_{\rm ph}$) in blue and on the PSB~2 ($\Delta_{\rm laser}=2\hbar\omega_{\rm ph}$) in red. (f) Gained displacement of the phonon state as a function of the initial phonon displacement. Approximations from Eq.~\eqref{eq:shift} and~\eqref{eq:shift_b} as solid and dashed lines, respectively and the full model as dots. Colors as in (e)}\label{fig:psbs_wigner}
\end{figure}

In particular, the form-stable Wigner function constitutes a good tool to interpret the scattering spectra studied below.
In order to start from the semiclassical limit we consider a rather large phonon amplitude of $u_0=10\gg 1$ and choose the coupling strength to $\gamma = 0.01\ll 1$ which is a typical value for the coupling between a QD exciton and LO phonons.\cite{reiter2011gene} In Fig.~\ref{fig:psbs_wigner} (a-d) we plot the Wigner function $W_x$ of the phonon state in the excited state of the TLS $\left|x\right>$ after one phonon period $t=t_0+T_{\rm ph}=t_0+2\pi/\omega_{\rm ph}$. For simplicity, we do not consider a radiative decay here, i.e., we let $\Gamma \to 0$ which includes the limiting case of an arbitrarily weak optical field $E_0\to 0$. We consider different laser energies increasing from PSB $-1$ ($\Delta_{\rm laser}=-\hbar\omega_{\rm ph}$) in (a) to PSB 2 ($\Delta_{\rm laser}=2\hbar\omega_{\rm ph}$) in (d). In the figures the lines crossing at $(U-u_0,\Pi)=(0,0)$ mark the maximum of the initial coherent phonon state. The maximum positions of the Wigner functions show us that for negative and vanishing laser detunings in (a) and (b) the phonon state remains unaffected by the light excitation. However, for positive detunings in (c) and (d) the maximum of the phonon distribution is shifted to larger displacements $U$. The reason is that the system generates additional phonons to reach an excited state occupation for positive detunings. To get an approximate quantification of this process we take a look at the energy balance derived in Appendix~\ref{sec:weak} Eq.~\eqref{eq:balance_app}, which relates the laser energy with the change of energy $\delta E$ in the coupled TLS-phonon system and the change of the occupation $\delta\left< \left|x\right>\!\left< x\right|\right>=\delta f$ via
\begin{align}\label{eq:balance_main}
	\hbar\overline{\Omega}+\Delta_{\mathrm{laser}}=\frac{\delta E}{\delta f}\,.
\end{align}
To fill this relation with more meaning, we first express the optically induced change of the state $\rho(t)$ with respect to the initial state $\rho_0$ as
\begin{subequations}\begin{align}
	\rho(t) &= \rho_0 + \delta \widetilde \rho (t)\,,\\
		\delta \widetilde \rho&= \left(\delta\widetilde\rho_{{\rm ph}, x} \left| x \right>\left< x \right| + \delta \widetilde \rho_{{\rm ph}, g}\left| g \right> \left< g \right|\right)+\text{off-diag.}\,,\\
	\rho_0 &= \left|g,\frac{u_0}{2}\right>\!\left<g,\frac{u_0}{2}\right|\,,
\end{align}\end{subequations}%
where
\begin{subequations}\begin{align}
		\delta \widetilde \rho_{{\rm ph}, x}(t) &= {\rm Tr}_{\rm TLS}\left[ \rho(t)\left |x \right>\!\left< x \right|\right] \,,\\
	\delta \widetilde \rho_{{\rm ph}, g}(t) &= {\rm Tr}_{\rm TLS}\left\{ \left[ \rho(t)-\rho_0 \right] \left |g \right>\!\left< g \right|\right\} \,,
\end{align}\end{subequations}%
and 'off-diag.' denotes terms that are non-diagonal in the subspace of the TLS.
We have marked the variations of the density matrix $\delta\widetilde\rho$ with a tilde to indicate that they do not have all the properties of a density matrix. Since the traces over $\rho(t)$ and $\rho_0$ are both unity, the trace over $\delta \widetilde \rho$ has to vanish. This implies
\begin{equation}\label{eq:delta_f}
	-{\rm Tr}(\delta\widetilde\rho_g)={\rm Tr}(\delta\widetilde\rho_x)={\rm Tr}\left[ \rho(t)\left |x \right>\!\left< x \right|\right]=\delta f\,,
\end{equation}
since $\left|x\right>$ is initially unoccupied. Now we can calculate the change of energy in the coupled TLS-phonon system as
\begin{align}
	\delta E&={\rm Tr}\left(H_0\rho\right)-{\rm Tr}\left(H_0\rho_0\right) \notag\\
		&={\rm Tr}\left(H_0\delta \widetilde \rho\right)\notag\\
		&={\rm Tr}\left(\delta \widetilde \rho_{{\rm ph}, g}\left< g \right|H_0\left |g \right>\right)+{\rm Tr}\left(\delta \widetilde \rho_{{\rm ph}, x}\left< x \right|H_0\left |x \right>\right)
\end{align}
with $H_0$ corresponding to the Hamiltonian from Eq.~\eqref{eq:H} in absence of optical driving. Using the properties
\begin{subequations}\begin{align}
	\left< g \right|H_0\left |g \right>&=\hbar\omega_\text{ph}b^{\dagger}b\,,\\
	\left< x \right|H_0\left |x \right>&=\hbar\omega_\text{ph}\tilde{b}^{\dagger}\tilde{b}+\hbar\overline{\Omega}\,,\\
	\tilde{b}&=b+\gamma \label{eq:b_tilde}
\end{align}\end{subequations}%
together with Eqs.~\eqref{eq:balance_main} and~\eqref{eq:delta_f}, we arrive at
\begin{subequations}\begin{align}
	\Delta_{\mathrm{laser}}&=\hbar\omega_{\rm ph}\left(-\delta n_{\mathrm{ph},g}+\delta n_{\mathrm{ph},x}\right)\,,\\
	\delta n_{\mathrm{ph},g}&=\frac{{\rm Tr}\left(\delta \widetilde \rho_{{\rm ph}, g}b^{\dagger}b\right)}{{\rm Tr}\left(\delta \widetilde \rho_{{\rm ph}, g}\right)}\,,\\
	\delta n_{\mathrm{ph},x}&=\frac{{\rm Tr}\left(\delta \widetilde \rho_{{\rm ph}, x}\tilde{b}^{\dagger}\tilde{b}\right)}{{\rm Tr}\left(\delta \widetilde \rho_{{\rm ph}, x}\right)}\,.
\end{align}\end{subequations}%
Here, $\tilde{b}$ denotes the phonon annihilation operator relative to the shifted phase space equilibrium. The $\delta n_{\rm{ph}, g/x}$ describe how the excess energy from the laser detuning is distributed among the phonons associated with the ground state $\left|g\right>$ and those associated with the excited state $\left| x\right>$, respectively. To make the connection with the change of the Wigner function, we have to consider
\begin{subequations}\begin{align}
	{\rm Tr}\left(\rho_0 H_0\right)={\rm Tr}\left(H_0\left|g,\frac{u_0}{2}\right>\!\left<g,\frac{u_0}{2}\right|\right)&=\hbar\omega_{\rm ph}\frac{u_0^2}{4}= E_0 ,\\
	{\rm Tr}\left(H_0\left|x,\frac{u_0}{2}-\gamma\right>\!\left<x,\frac{u_0}{2}-\gamma\right|\right)&=E_0+\hbar\overline{\Omega}\,.
\end{align}\end{subequations}%
This implies that the initial phonon distribution associated with $\left| g\right>$ and its shifted counterpart in $\left| x\right>$ have the same total energy. We now define the energy change of the phonons associated with the deviation from the original distributions in phase space as
\begin{equation}\label{eq:delta_E}
	\delta E_{\mathrm{ph},g/x}=\hbar\omega_{\rm ph}\delta n_{\mathrm{ph},g/x}-E_0\,,
\end{equation}%
which leads to
\begin{equation}\label{eq:balance_final}
	\Delta_{\mathrm{laser}}=\delta E_{\mathrm{ph},x}-\delta E_{\mathrm{ph},g}=\delta E_{\mathrm{ph}}\,.
\end{equation}%

We show the distribution of the energy on the subsystems in Fig~\ref{fig:psbs_wigner}(e), where we plot $-\delta E_{{\rm ph}, g}$ and $\delta E_{{\rm ph}, x}$ for different initial displacements $u_0$ as dots and squares, respectively. The blue symbols represent an optical excitation on PSB~1 ($\Delta_{\rm laser}=\hbar\omega_{\rm ph}$) and the red ones on PSB~2 ($\Delta_{\rm laser}=2\hbar\omega_{\rm ph}$). In Appendix~\ref{sec:ground} we derive the ground state's generating function that is used to calculate $\delta E_{{\rm ph},g}$. We observe that for a vanishing $u_0$, i.e., the initial phonon vacuum, the energy of the ground state phonons does not change, i.e., $\delta E_{\mathrm{ph},g}=0$, while the excited state phonons gain the detuning energy $\delta E_{{\rm ph},x}=\Delta_{\rm laser}$. However, for non-vanishing coherent amplitudes ($u_0\neq 0$) $-\delta E_{{\rm ph}, g}$ additionally becomes negative. This excess energy is brought into the excited state's phonon system on top of the previously described energy gain resulting in $\delta E_{{\rm ph},x}=\left|\delta E_{{\rm ph},g}\right|+\Delta_{\rm laser}$. In the case of a far displaced coherent state with $u_0\gg1$, the ground state phonons lose the energy of the detuning $\delta E_{{\rm ph},g}=\Delta_{\rm laser}$, while consequently the excited state's phonons gain twice the detuning energy 
\begin{equation}\label{eq:2Delta}
	\delta E_{{\rm ph},x}=2\Delta_{\rm laser}\,.
\end{equation}%
To confirm that the entire energy balance in Eq.~\eqref{eq:balance_final} holds, in Fig.~\ref{fig:psbs_wigner}(e) we plot the entire energy gain $\delta E_{\mathrm{ph}}$ in the phonon system as crosses. We find that these always agree with the respective laser detunings, marked by the blue and red dashed line.

In a final step we link the energy balance to the increase of the displacement for the case of $u_0\gg 1$ (see Fig.~\ref{fig:psbs_wigner}(c,\,d)). In this case with Eqs.~\eqref{eq:delta_E} and \eqref{eq:2Delta} we get
\begin{align}
	2\Delta_{\rm laser}&= \hbar\omega_\text{ph} \frac{{\rm Tr}\left(\delta \widetilde \rho_{{\rm ph}, x}\tilde{b}^{\dagger}\tilde{b}\right)}{{\rm Tr}\left(\delta \widetilde \rho_{{\rm ph}, x}\right)}-E_0\notag\\
	&= \hbar\omega_{\rm ph} \left<\tilde{b}^{\dagger}\tilde{b}\right>_x-E_0\,,
\end{align}
where the average $\left<\dots\right>_x$ is with respect to $\left<x\right|\rho(t)\left|x\right>=\delta \widetilde \rho_{{\rm ph}, x}$.
Now assuming that the phonons remain in a coherent state, this relation can be rewritten as
\begin{equation}
2\Delta_{\mathrm{laser}}\approx\hbar\omega_{\rm ph} \left|\left<b+\gamma\right>_x\right|^2-E_0
\end{equation}
Noting that the phonon states in Fig.~\ref{fig:psbs_wigner} are evaluated at full periods of the harmonic dynamics the coherent amplitudes are real and we can identify
\begin{align}
	\left<u\right>_x =2\left<b\right>_x\,,
\end{align}
and introduce the displacement shift
\begin{align}
	\delta\!\left<u\right> = \left<u\right>_x + 2\gamma - u_0\,,
\end{align}
which compensates the equilibrium shift by $-2\gamma$. Then we arrive at
\begin{align}
	2\Delta_{\rm laser}&\approx \frac14 \hbar\omega_\text{ph} \left[ \left(\delta\!\left<u\right>\right)^2 + 2u_0 \delta\!\left<u\right>\right]\,.
\end{align}%
From this we can finally retrieve the gained displacement of the Wigner function to
\begin{subequations}\label{eq:shift_all}
\begin{align}
	\delta\left< u\right> &= \sqrt{u_0^2 + 8\frac{ \Delta_{\rm laser}}{\hbar\omega_\text{ph}}} - u_0\label{eq:shift}\\
	 &\approx \frac{4 \Delta_{\rm laser} }{\hbar\omega_{\rm ph} u_0 }\, .\label{eq:shift_b}
\end{align}\end{subequations}%
In the second line we have taken the limit of a large initial displacement $u_0^2\gg 8\Delta_{\rm laser}/(\hbar\omega_{\rm ph})$. This shows that the gained displacement becomes very small for large initial displacements $u_0$. We will refer to these approximations as semiclassical limit.

To test the approximations in Eqs.~\eqref{eq:shift_all} in Fig.~\ref{fig:psbs_wigner}(f) we plot the gained displacement of the Wigner function, i.e., the expectation value difference $\delta\left<u\right>$, as a function of the initial displacement $u_0$ for different laser detunings $\Delta_{\rm laser}$. The results of the full model are shown as dots and the approximations from Eqs.~\eqref{eq:shift} and \eqref{eq:shift_b} as solid and dashed lines, respectively, where the different colors represent the detunings $\Delta_{\rm laser} = \hbar\omega_{\rm ph}$ (PSB 1, blue) and $\Delta_{\rm laser} = 2\hbar\omega_{\rm ph}$ (PSB 2, red). The approximations reflect two aspects which we find in the simulations of the full model. For large initial displacements $u_0$, i.e., in the semiclassical limit, the gained displacement $\delta\!\left< u \right>$ vanishes because already a small additional displacement compensates the excess energy from the detuning. In addition, we see that the gained displacement is larger for larger laser detunings, which is expected because more energy has to be compensated by phonon generation to reach an excited state occupation. For small $u_0$ however, the assumption of a coherent phonon state is invalid.  From the limiting case of an initial vacuum state ($u_0=0$) we know that for small initial displacements, the energy is not converted into an additional displacement but leads to a change of the phonon Fock state without changing the displacement apart from the shift of the equilibrium. This is why the shifts obtained from the full model vanish in this limit and therefore deviate qualitatively from the approximations. In the particular case $u_0=0$, the shift $\delta\!\left<u\right>$ even vanishes because the state becomes a Fock state in the excited state's equilibrium position, i.e., at $\left<u\right>= - 2\gamma$.~\cite{groll2021controlling}

\subsection{Scattering spectra in the classical-to-quantum transition regime}
In the next step we systematically approach the quantum regime by calculating the light scattering spectra for decreasing initial displacements of coherent states. At the same time we study the modification of the Wigner function, which allows to gain additional information to understand the optical spectra.

\subsubsection{Resonant excitation}\label{sec:res}
\begin{figure}[!t]
\includegraphics[width = \columnwidth]{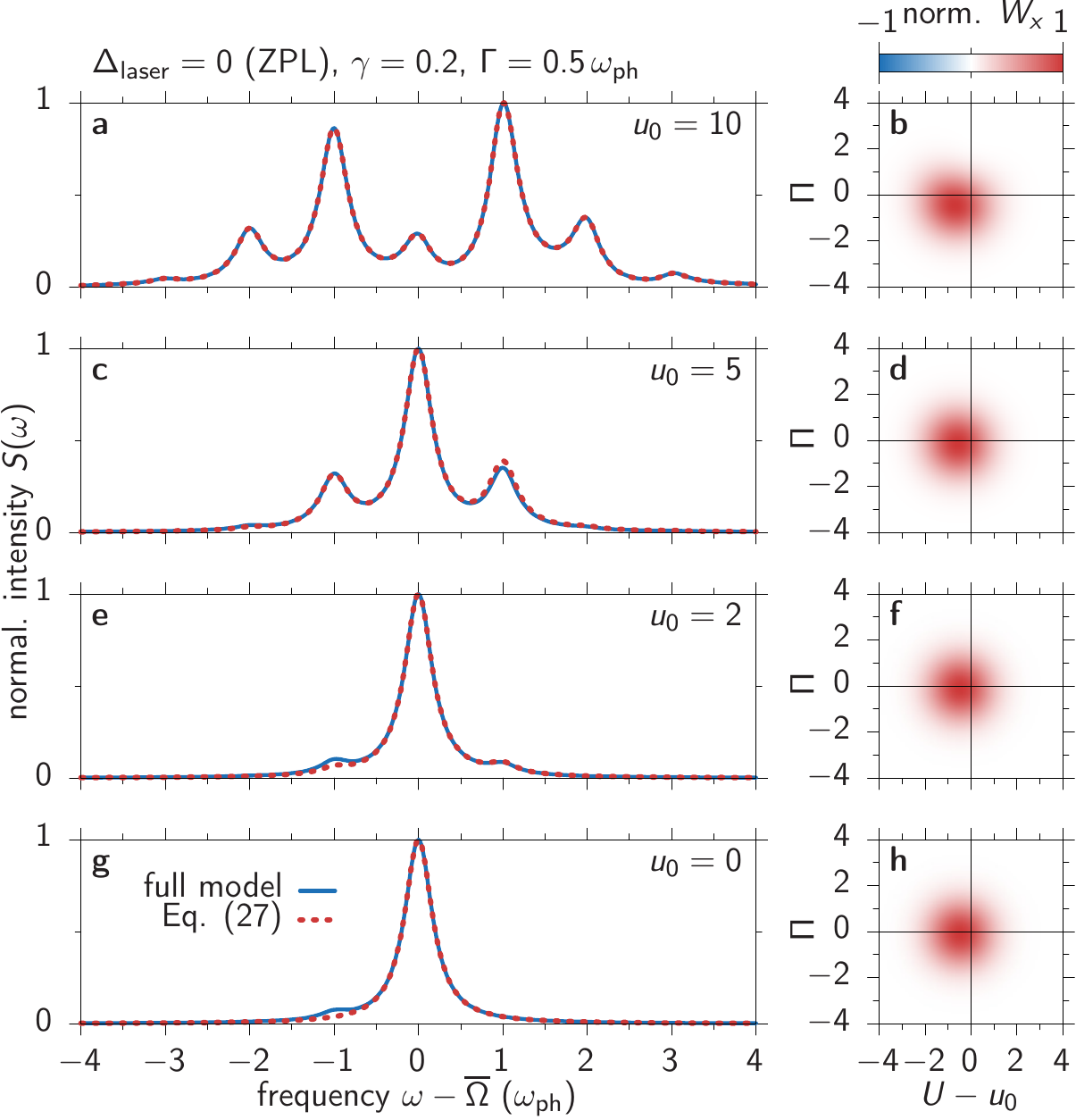}
\caption{Resonant classical-to-quantum transition. The laser energy is in resonance with the TLS transition (ZPL) and the phonon coupling strength is $\gamma=0.2$. (a), (c), (e), (g) Normalized scattering spectra with the full model as solid and the approximation employing Eq.~\eqref{eq:classical} as dashed lines. (b), (d), (f), (h) Normalized Wigner functions $W_x$ at full phonon periods. The initial displacement $u_0$ decreases from top to bottom as given in the plots. The spectra are additionally broadened by $\eta=0.2\omega_{\rm ph}$ to include a non-vanishing detection resolution.}\label{fig:spec_zpl}
\end{figure}

We start with $u_0=10$, a laser excitation in resonance with the ZPL ($\Delta_{\rm laser}=0$), and we include a decay of $\left|x\right>$ with the rate $\Gamma = 0.5\omega_\text{ph}$. We consider here the time-integrated spectrum in Eq.~\eqref{eq:S_int} after the Wigner function $W_x$ has become form-stable, i.e., for $T_0 \gg t_0$, and the period $T=T_{\rm ph}$. The simulated scattering spectrum is shown in Fig.~\ref{fig:spec_zpl} (a), where the solid blue line represents the full model and the dashed red line the approximated spectrum in the semiclassical limit employing Eq.~\eqref{eq:classical}. The spectral lines are broadened with a linewidth of $\eta = 0.2\omega_\text{ph}$. The combination of coupling strength $\gamma=0.2$ and initial detuning $u_0=10$ represents such a strong phonon impact that the resonantly scattered light in the ZPL only has a minor contribution, while the first PSBs ($\pm1$) dominate the spectrum. Such strong coupling constants can for example be found for core-shell QDs\cite{reiter2011gene} or color centers in hexagonal boron nitride.\cite{groll2021controlling} The large displacement field might be achieved in SAW wave guide\cite{vogele2020quantum} or resonator systems.\cite{whiteley2019spin} Both simulations agree perfectly, which demonstrates that such an initial displacement can be accurately described in the semiclassical limit. This is further confirmed by the Wigner function $W_x$ of the final phonon state in Fig.~\ref{fig:spec_zpl} (b), which is almost perfectly symmetric, i.e., it remains coherent during the light scattering.

To reach the ultimate quantum state, represented by the phononic vacuum, we decrease the initial displacement to $u_0=5$ and 2 and retrieve the scattering spectra in Figs.~\ref{fig:spec_zpl} (c) and (e), respectively. Here, we already find slight differences between the full model (solid blue curves) and the approximation in the semiclassical limit (dashed red curves). Still, the Wigner functions in Fig.~\ref{fig:spec_zpl} (d) and (f) remain Gaussian and therefore coherent.

Finally, for the initial phonon vacuum state with $u_0=0$ in Fig.~\ref{fig:spec_zpl} (g) the semiclassical approximation (dashed line) only provides the elastically scattered ZPL at $\omega=0$. The semiclassical limit shows here no impact of the phonons at all because in this approximation the TLS cannot act as a spontaneous source of phonons: On the one hand, phonon absorption and stimulated emission that could lead to PSBs are only possible for a non-vanishing amplitude $u_0$. The full simulation on the other hand results in an additional small PSB $-1$ at $\omega=-\omega_{\rm ph}$ which describes the possibility of spontaneous phonon generation even for an initial phonon vacuum state. The respective Wigner function in Fig.~\ref{fig:spec_zpl} (h) still keeps its coherent shape and only experiences a slight shift in phase space, due to the shift of the excited state's equilibrium displacement by $-2\gamma=-0.4$.\cite{hahn2019infl}

\subsubsection{Detuned excitation}
\begin{figure}[b]
\includegraphics[width = \columnwidth]{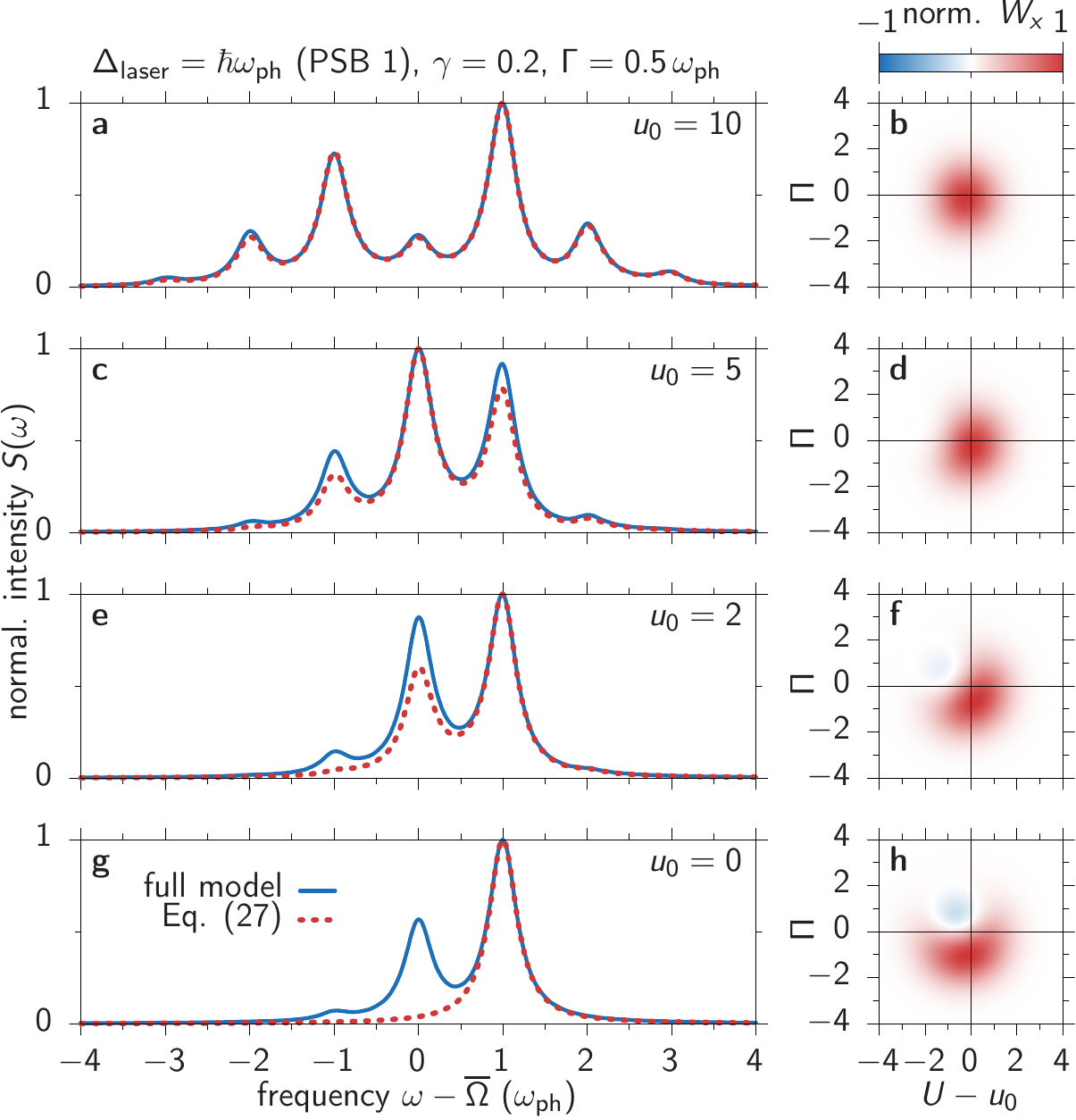}
\caption{Classical-to-quantum transition. The laser energy is in resonance with the first phonon-assisted transition (PSB 1) and the phonon coupling strength is $\gamma=0.2$. (a), (c), (e), (g) Scattering spectra with the full model as solid and the approximation employing Eq.~\eqref{eq:classical} as dashed lines. (b), (d), (f), (h) Normalized Wigner functions $W_x$ at full phonon periods. The initial displacement $u_0$ decreases from top to bottom as given in the plots. The spectra are additionally broadened by $\eta = 0.2\omega_{\rm ph}$ to include a non-vanishing detection resolution.}\label{fig:spec_psb}
\end{figure}

In Sec.~\ref{sec:phonons} we have already seen that a detuned excitation can significantly affect the phonon state. In Fig.~\ref{fig:spec_psb} we perform the same study as in Sec.~\ref{sec:res} but choose the laser energy in resonance with the PSB 1, i.e., $\Delta_{\rm laser}=\omega_{\rm ph}$. Here, the case for the largest initial displacement with $u_0=10$ in Figs.~\ref{fig:spec_psb} (a) and (b) shows the expected result of being in the semiclassical limit. The full model of the scattering spectrum (solid blue line) in (a) and the semiclassical approximation (dashed red line) agree well and the Wigner function in (b) remains Gaussian, i.e., coherent.

The differences between the full model and the semiclassical approximation become already obvious for $u_0=5$ in Figs.~\ref{fig:spec_psb} (c) and (d). The peak intensities deviate slightly between the full model and the semiclassical approximation. Also the Wigner function already looses its symmetric shape and becomes slightly squeezed.

While for an initial displacement of $u_0=2$ in Figs.~\ref{fig:spec_psb} (e) and (f) the differences in the scattering spectrum increase but do not qualitatively change, the Wigner function exhibits a new interesting feature. We find negativities in the distribution, which are a clear indication of quantum behavior.\cite{kenfack2004negativity} The curved positive region bending around the circular negative part is typical for Fock state superpositions of the form $\alpha \left| \tilde{0}\right> + \beta\left| \tilde{1}\right>$, which are known to exhibit squeezing.~\cite{drummond2013quantum, wigger2016quan} Here, the phonon states are given as Fock states in the shifted phase space of the excited state of the TLS with $\tilde{b}^\dagger\tilde{b}\left|\tilde{n}\right>=\tilde{n}\left|\tilde{n}\right>$ ($\tilde b$ from Eq.~\eqref{eq:b_tilde}).

When considering the ultimate quantum limit of an initial vacuum state with $u_0=0$ the scattering spectrum in Fig.~\ref{fig:spec_psb} (g) shows the known features: the semiclassical approximation (dashed red line) only provides the elastically scattered laser light at $\omega=\omega_{\rm ph}$ as explained before, while the full model (solid blue line) shows two additional PSBs below the laser energy. These peaks again represent spontaneous phonon generation processes. The respective Wigner function in Fig.~\ref{fig:spec_psb}~(h) still exhibits the quantum superposition characteristics discussed before. However, it does not represent the pure shifted Fock state $\left|\tilde 1\right>$ as would be expected in the limiting case of an optical excitation that is much faster than the excited state decay.\cite{groll2021controlling} Here, the interplay between constant optical driving and decay results in the combination of $\left|\tilde 0\right>$ and $\left|\tilde 1\right>$.

\begin{figure}[t]
\includegraphics[width = \columnwidth]{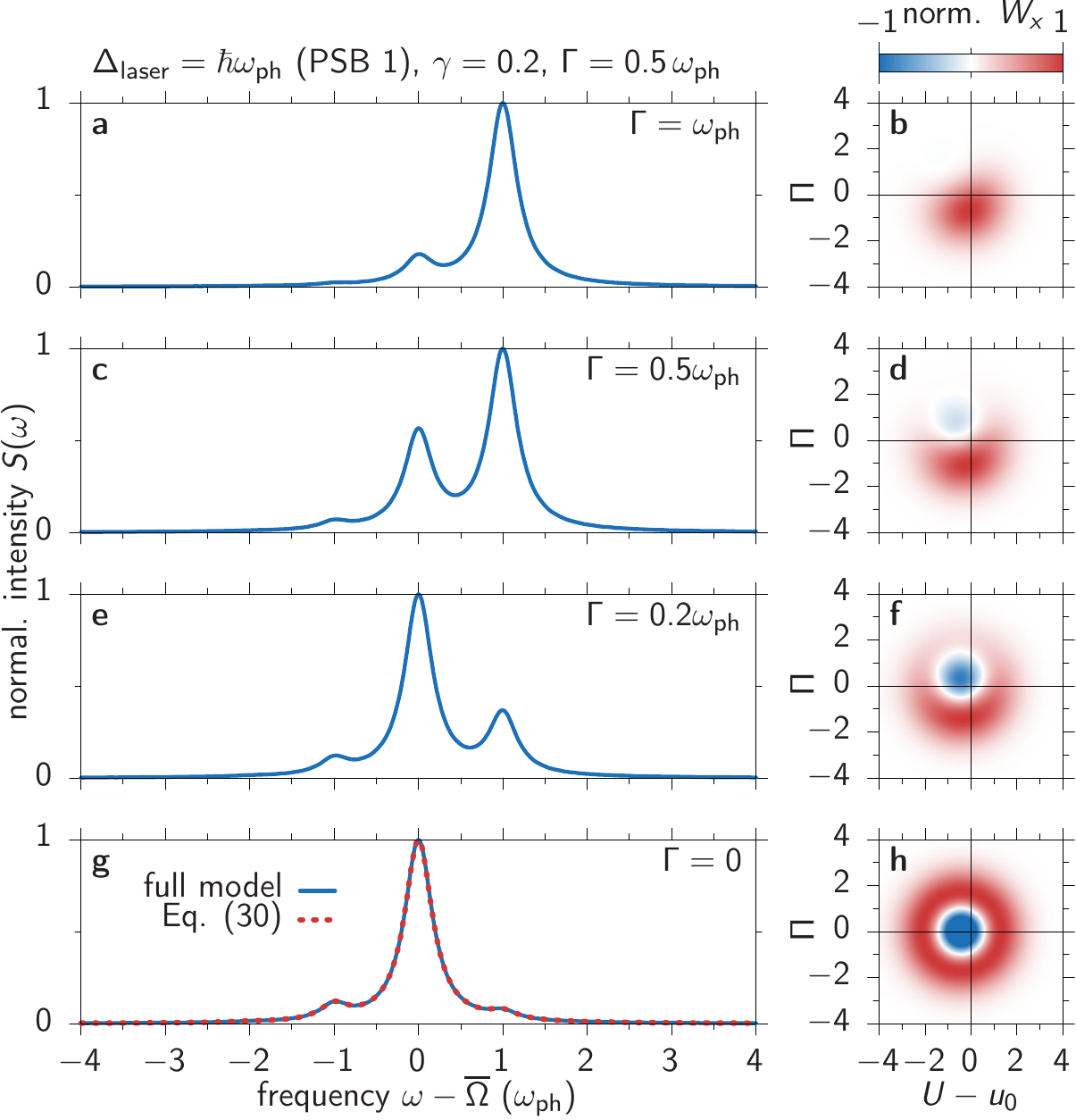}
\caption{Impact of the decay rate $\Gamma$. The laser energy is in resonance with the first phonon-assisted transition (PSB 1) and the phonon coupling strength is $\gamma=0.2$. (a), (c), (e), (g) Scattering spectra with the full model. For $\Gamma = 0$, the spectrum from Eq.~\eqref{eq:RF_vac} is given by the dotted red line. (b), (d), (f), (h) Normalized Wigner functions $W_x$ at full phonon periods. The spectra are additionally broadened by $\eta = 0.2\omega_{\rm ph}$ to include a non-vanishing detection resolution.}\label{fig:spec_psb_Gamma}
\end{figure}

Finally, we analyze the impact of the TLS decay rate $\Gamma$ on the scattering spectra and the phonon state. They are depicted for different decay rates in Fig.~\ref{fig:spec_psb_Gamma} with $\Gamma$ decreasing from top to bottom as given in the plot. For a decay rate on the order of the phonon frequency $\Gamma=\omega_{\rm ph}$ and $0.5\omega_{\rm ph}$ as shown in Fig.~\ref{fig:spec_psb_Gamma} (a) and (c), respectively, we find that the most pronounced spectral peak appears at the laser frequency $\omega-\overline{\Omega}=\Delta_\text{laser} = \hbar \omega_\text{ph}$. But we already see that the phonon-assisted peaks at frequencies smaller than the laser gain weight when the decay rate is reduced. The corresponding Wigner functions in (b) and (d) remain predominantly positive which shows that the wave function is to a large degree given by the vacuum state~$\left|\tilde 0\right>$.

In contrast, for a smaller decay rate of $\Gamma=0.2\omega_{\rm ph}$ the Wigner function in Fig.~\ref{fig:spec_psb_Gamma}~(f) exhibits the characteristic ring structure of the Fock state $\left|\tilde 1\right>$. And also the scattering spectrum is now dominated by the peak at $\omega=\overline{\Omega}$. In the limit of a vanishing decay $\Gamma=0$, the Wigner function in Fig.~\ref{fig:spec_psb_Gamma} (h) has a perfectly circular symmetric shape around the shifted equilibrium position of $\left|x\right>$. This already indicates that the system is brought into the displaced Fock state $\left|x,\tilde 1\right>$. The corresponding scattering spectrum in (f) is almost entirely given by the peak at the TLS transition with $\omega=\overline{\Omega}$ and only minor contributions at the neighboring phonon frequencies appear. This result is further confirmed by the analytical vacuum approximation employing Eq.~\eqref{eq:RF_vac} depicted as red dotted line which agrees perfectly with the full model. 

We can again understand the transition from fast to slow decay rates in terms of the energy balance in the system. In the limit of a slow TLS decay the excess laser energy provided by the detuning $\Delta_{\rm laser}=n\hbar\omega_{\rm ph}$ is used to generate the corresponding displaced phononic Fock state $\left|\tilde n\right>$ and the light is emitted from the bare TLS transition at $\omega=\overline{\Omega}$ and its corresponding PSBs. In the limit of a rapid TLS decay the phononic system does not have enough time to generate a different quantum state before the TLS is re-emitting the light in the decay process. Therefore, the majority of the light is scattered elastically with the frequency $\omega=\overline{\Omega}+\Delta_{\rm laser}$.

\section{conclusions}
In this proof-of-concept study we have developed a theoretical method to calculate the resonance fluorescence spectrum of a single TLS in the presence of an arbitrary phonon state of a mechanical single mode system. The treatment is based on generating functions that allow us to calculate the density matrix dynamics without approximations in the phonon coupling. In the limit of low optical excitation amplitudes we have developed an expression for the two-time correlation function that is used to determine the optical scattering spectrum. In addition, from the generating functions we have access to the modification of the phonon state by the optical driving of the TLS as displayed by Wigner functions.

To test this approach, we have performed a study of the classical-to-quantum regime of the considered phonon state. For an initial coherent state with a large amplitude we have shown that our quantum model exhibits the same results as the semiclassical treatment, which was tested in previous studies.\cite{weiss2021opto, wigger2021reso} We showed that the initial phonon state remains unaffected by the optical scattering in this regime. By systematically reducing the amplitude of the coherent state we crossed the boundary to the quantum regime and found that both, the optical scattering spectrum and the phonon state exhibit significant changes. We further spotted that for detuned excitations and TLS decay rates significantly faster than the phonon period the scattering spectrum is dominated by elastic scattering. The situation changed significantly when we considered a slow TLS decay where the phonons were brought into a higher Fock state and the light scattering was dominated by the TLS transition and the corresponding PSBs.

After this initial study of the impact of phononic quantum states on the resonance fluorescence spectra of single-photon emitters it would be interesting to explore the impact of more involved quantum states on the optical spectrum. Typical examples might be superpositions like Schr\"odinger cat states or squeezed states and it could be interesting to examine the possibilities to read initial phononic quantum properties from the features of the optical scattering spectrum.

\appendix

\section{Form-stable state of Wigner functions}\label{sec:stable}
When calculating the form-stable state of the generating functions in Eq.~\eqref{eq:c2}, i.e., $C^{(2)}(\alpha , t\to\infty)$, under optical excitation with the detuning $\Delta_\text{laser} = m\hbar\omega_\text{ph}$, $m\in\mathbb Z$, we are confronted with two integrals over the product of a time-periodic function $f(t) = f(t+T_{\rm ph})$ and an exponential decay. Each of these calculations can be reduced to the integral over one period via the geometric series
\begin{align}
	\int\limits_{t_0}^{\ \infty} f(t') e^{-\Gamma t'}dt' = \frac{1}{1-e^{-\Gamma T_{\rm ph}}} \int\limits_{t_0}^{\ t_0+ T_{\rm ph}} f(t') e^{-\Gamma t'}dt'.
\end{align}
Therefore, to determine the Wigner function of the phonon state in Eq.~\eqref{eq:Wigner} in the form-stable state it is sufficient to determine $C(\alpha,t)$ after one phonon period at $t=t_0+T_{\rm ph}$. Note, that we have assumed that the upper limit of the integration is given by $t_0+nT_{\rm ph}\to \infty$, $n\in\mathbb N$. When not considering an integration until a full phonon period, the Wigner function has the same shape but is rotated differently with respect to the shifted equilibrium in phase space, i.e., the fixed point as described in the main text.

\section{Derivation of the energy balance for weak optical driving}\label{sec:weak}
To derive the energy balance from Eq.~\eqref{eq:balance_main} which relates the detuning of the laser with the change of phonon statistics, we start by considering an auxiliary quantum optical system with a single photon mode. Note, that our model is the semi-classical limit of this system. In practice, instead of the external classical cw-driving, we imagine a single photon mode coupled to the TLS. In such systems,\cite{groll2020four} the number of photons $N_{\mathrm{phot}}$ and the occupation $f$ of the TLS are related by
\begin{align}
	&N_{\mathrm{photon}}+f=\text{const.}\notag\\
	\Rightarrow &\delta N_{\mathrm{photon}}=-\delta f\,.
\end{align}%
Any change (denoted by $\delta$) in the number of photons is thus compensated by a change of equal size in the occupation of the TLS. Since we consider weak optical driving in the main text, in our auxiliary system the TLS and photons are coupled weakly. Therefore, the total energy $E_{\mathrm{total}}$ is approximately given by a sum of the photon energy $E_{\mathrm{photon}}$ and the energy $E_{\mathrm{system}}$ of the coupled TLS-phonon system
\begin{equation}
	E_{\mathrm{total}}=E_{\mathrm{photon}}+E_{\mathrm{system}}\,.
\end{equation}%
Contrary to our model in the main text which is driven externally, the auxiliary system is closed as the photons are included. Therefore the total energy is constant, yielding the following relation for the changes of $E_{\mathrm{photon}}$ and $E_{\mathrm{system}}$
\begin{align}
	0&=\delta E_{\mathrm{photon}}+\delta E_{\mathrm{system}}\notag\\
		&=\hbar \omega_{\mathrm{photon}}\delta N_{\mathrm{photon}}+\delta E_{\rm system}\notag\\
		&=-\hbar\omega_{\mathrm{photon}}\delta f+\delta E_{\mathrm{system}}\,.
\end{align}%
Here we introduced the energy $\hbar\omega_{\mathrm{photon}}$ of the photons, which relates to the carrier frequency of the laser in our semiclassical model via
\begin{equation}
	\hbar\omega_{\mathrm{photon}}=\hbar\overline{\Omega}+\Delta_{\mathrm{laser}}\,.
\end{equation}%
With this we arrive at
\begin{equation}\label{eq:balance_app}
\hbar\overline{\Omega}+\Delta_{\mathrm{laser}}=\frac{\delta E_{\mathrm{system}}}{\delta f}\,,
\end{equation}
relating the carrier frequency of the laser to the change of energy in the coupled TLS-phonon system. Note that for simplicity we drop the index 'system' in the main text.

\section{Ground state phonons}\label{sec:ground}
As the light field transports occupation from the ground state into the excited state it changes the phonons of both subsystems $\left| g\right>$ and $\left| x\right>$. Analogous to the calculation of the excited state, we obtain the ground state's generating function by integrating Eq.~\ref{eq:G}, inserting ${\overline Y}^{(1)}$ and transforming it back with Eq.~\ref{eq:bG}. With this procedure we obtain
\begin{align}
	&G^{(2)}(\alpha, t) =\notag\\
		& -\int\limits_0^t\int\limits_0^{t'}dt'dt'' {\overline E}^*(t'){\overline E}(t'')\notag\\
			&\qquad \times \exp\left\{ \gamma \left[\alpha e^{i\omega_{\rm ph}(t'-t)}+\gamma\right]\left[e^{i\omega_{\rm ph}(t''-t')}-1\right]\right\}\notag\\
			&\qquad\times {\overline G}^{(0)}\left( \alpha e^{-i\omega_{\rm ph}t} + \gamma e^{-i \omega_{\rm ph} t'} - \gamma e^{-i\omega_{\rm ph}t''}\right) \notag\\
		&+\int\limits_0^t\int\limits_0^{t'}dt'dt''{\overline E}^*(t''){\overline E}(t')\notag\\
			&\qquad\times \exp\left\{ \gamma \left[-\alpha^* e^{-i\omega_{\rm ph}(t'-t)}+\gamma\right]\left[e^{-i\omega_{\rm ph}(t''-t')}-1 \right]\right\} \notag\\
			&\qquad\times{\overline G}^{(0)}\left( \alpha e^{-i\omega_{\rm ph}t} - \gamma e^{-i \omega_{\rm ph} t'} + \gamma e^{-i\omega_{\rm ph}t''}\right) \,.
\end{align}%

\section*{acknowledgments}
T.H. thanks the German Academic Exchange Service (DAAD) for financial support (No.~57504619). D.W. thanks the Polish National Agency for Academic Exchange (NAWA) for financial support within the ULAM program (No.~PPN/ULM/2019/1/00064). T.H., D.G., T.K., P.M., and D.W. acknowledge support from NAWA under the APM grant No.~PPI/APM/2019/1/00085.

\section*{Data available}
The data that support the findings of this study are available from the corresponding author upon reasonable request.

\section*{Author Declarations}
The authors have no conflicts to disclose.


%

\end{document}